# Star formation in a galactic outflow

R. Maiolino[1,2], H.R. Russell[3], A.C. Fabian[3], S. Carniani[1,2], R. Gallagher[1,2], S. Cazzoli[4], S. Arribas[4], F. Belfiore[1,2], E. Bellocchi[4], L. Colina[4], G. Cresci[5], W. Ishibashi[6], A. Marconi[5,7], F. Mannucci[5], E. Oliva[5], E. Sturm[8]

[1] Cavendish Laboratory, University of Cambridge, 19 J. J. Thomson Ave., Cambridge CB3 0HE, UK
[2] Kavli Institute for Cosmology, University of Cambridge, Madingley Road, Cambridge CB3 0HA, UK
[3] Institute of Astronomy, Madingley Road, Cambridge CB3 0HA, UK
[4] CSIC - Departamento de Astrofisica-Centro de Astrobiologia (CSIC-INTA), Torrejon de Ardoz, Madrid, Spain
[5] INAF - Osservatorio Astrofisico di Arcetri, Largo E. Fermi 5, 20125, Firenze, Italy
[6] Physik-Institut, Universität Zürich, Winterthurerstrasse 190, 8057 Zürich, Switzerland
[7] Dipartimento di Fisica e Astronomia, Università di Firenze, Via G. Sansone 1, 50019, Sesto Fiorentino (Firenze), Italy
[8] Max-Planck-Institut für Extraterrestrische Physik, Giessenbachstr., D-85748 Garching, Germany

**Recent observations have revealed massive galactic molecular outflows[1-3] that may have physical conditions (high gas densities[4-6]) required to form stars. Indeed, several recent models predict that such massive galactic outflows may ignite star formation within the outflow itself[7-11]. This star-formation mode, in which stars form with high radial velocities, could contribute to the morphological evolution of galaxies[12], to the evolution in size and velocity dispersion of the spheroidal component of galaxies[11,13], and would contribute to the population of high-velocity stars, which could even escape the galaxy[13]. Such star formation could provide *in-situ* chemical enrichment of the circumgalactic and intergalactic medium (through supernova explosions of young stars on large orbits), and some models also predict that it may contribute substantially to the global star formation rate observed in distant galaxies[9]. Although there exists observational evidence for star formation triggered by outflows or jets into their host galaxy, as a consequence of gas compression, evidence for star formation occurring *within* galactic outflows is still missing. Here we report new spectroscopic observations that unambiguously reveal star formation occurring in a galactic outflow at a redshift of 0.0448. The inferred star formation rate in the outflow is larger than 15 $M_\odot$/yr. Star formation may also be occurring in other galactic outflows, but may have been missed by previous observations owing to the lack of adequate diagnostics[14,15].**

IRAS F23128-5919 is a merging system (Fig.1a), in which the southern nucleus hosts an obscured active nucleus (AGN), detected in the X-rays[16]. Past observations had already revealed a prominent outflow developing from the southern nucleus[14,15,17-19], driven by the nuclear starburst, or by the AGN, or both. We analyzed archival Very Large Telescope (VLT) spectroscopic observations, obtained with the MUSE instrument, of the optical nebular lines to better characterize the outflow. The nebular emission line profiles can be clearly separated into a narrow component, associated with the interstellar medium in the two galactic disks, and a very broad (Full Width Half Maximum, FWHM~600-1,000 km/s), predominantly blueshifted component tracing the outflow. The velocity field, velocity dispersion, surface brightness maps of the two components are shown in Fig. 1a. The narrow component (bottom row in Fig. 1a) is probably tracing the bulk of the dynamics of the two merging disks. The outflow traced by the broad blueshifted component (top row in Fig. 1a) of the nebular lines extends towards the East of the southern nucleus for about 7-9 kpc (8"-10"), beyond the optical galactic disk. A receding counter-outflow is also observed in the opposite direction, i.e. towards the West (although weaker, owing to extinction by the galactic disk).

We have observed the central and eastern outflows with the X-shooter spectrograph at the VLT, which enabled the detection of spectral diagnostics over the entire spectral range from 0.35μm to 2.5μm, and with a spectral resolution and sensitivity higher than the MUSE data. The slit location (box) is shown in Fig.1a. An example of a spectrum extracted from the central region is shown in Figs.1b, zoomed around some of the relevant emission lines. The high spectral resolution and high sensitivity of the X-shooter spectra reveal that at least two or three Gaussian components are needed to properly reproduce the asymmetric profile of the broad component of the nebular emission lines, in addition to the narrow component tracing the interstellar medium in the host galaxy (see Methods). We will however show that the results do not depend critically on the Gaussian decomposition.

Figure 2a shows the location of the various emission line components tracing the outflow onto the so-called 'BPT diagnostic diagrams', which are widely used to obtain an approximate discrimination between different excitation mechanisms[20]. Most components describing the outflow (broad blueshifted components) are located in the region of the diagram populated by star forming galaxies and HII regions, with only a few exceptions mostly restricted to the diagram involving the [NII] line. However, this diagnostic has potential problems associated with the nitrogen abundance[21,22] and, in this specific target, the [NII] line is affected by telluric absorption, whose correction introduces additional uncertainties.

The finding that nebular components tracing the outflow are located in the star forming locus of the BPT diagrams already provides a first indication that star formation is probably occurring in the outflow. However, the BPT diagrams do not necessarily provide an unambiguous classification. For instance, there are excitation models that predict that some shocks could produce line ratios in the 'star forming' region of the BPT diagrams[23].

Fortunately, the X-shooter wide spectral range provides a wealth of additional diagnostics of the excitation mechanism. Shock/AGN excitation can be cleanly distinguished from excitation by young stars in the [FeII]1.64μm/Brγ versus $H_2$(1-0)S(1)2.12μm/Brγ diagram[24], as shown in Fig.2b (see Methods for details). Blue symbols indicate the line ratios of the components tracing the central part of the outflow in IRAS23128-5919, indicating that the gas excitation in the outflow is consistent with star formation and not consistent with other excitation mechanisms, such as shocks or AGNs. The [FeII]1.25μm/Paβ and [PII]1.18μm/Paβ diagram is also a good discriminator of the excitation mechanism[25], as illustrated in Fig.2c. The outflow in IRAS23128-5919 is inconsistent either with shock excitation or with AGN photoionization. The outflow line ratios are consistent with those observed in star-forming galaxies.

As illustrated in Fig.1b, the spectrum of IRAS23128-5919 does not show evidence for coronal lines, which are generally associated with powerful AGNs. This further supports the absence of significant AGN contribution to the gas excitation (see Methods for further discussion). Therefore, neither shocks nor the AGN can account for the excitation of the gas in the outflow, whereas all diagnostics are consistent with excitation by star formation.

The presence of young stars is clearly revealed from the ultraviolet Hubble Space Telescope (HST) images (Fig.1a). However, from imaging alone it is not possible to disentangle putative young stars in the outflow from young stars in the galaxy discs, whose ultraviolet radiation field can potentially ionize the gas in the outflow (externally) and produce the line ratios observed in the outflow.

However, an external source of stellar ionization should result in an ionization parameter (defined as the ratio between ionizing photon flux and gas density; when divided by the speed of light it is the adimensional ionization parameter U) much lower than typically observed in star-forming galaxies. Yet, despite the gas density in the outflow (~600-1,500 cm$^{-3}$, as inferred from the [SII] doublet) being higher than in the host galaxy and than that typically seen in star-forming galaxies, the ionization parameter of the gas in the outflow, as inferred from the [OIII]5007/[OII]3727 line ratio, is not lower than observed in normal star-forming regions, and it is in fact slightly larger, as illustrated in Fig.3. This result strongly argues in favor of *in-situ* stellar photoionization, by young stars within the outflow.

Even better evidence for stars formed in the outflow is the direct detection of a young stellar population with kinematical fingerprints of formation inside the outflow. The stellar continuum is detected in the X-shooter spectra out to about 5 kpc from the nucleus, and its shape and stellar features do indeed indicate the presence of a young stellar population younger than a few tens of million years (Myr). Determining the kinematics of such a stellar population is difficult, since most stellar features (especially the ones associated with young stars) are heavily contaminated by the strong nebular emission lines. However, we have recovered the kinematics of the young stellar populations through our spectral fitting of the optical spectrum (Fig.4a and Methods), which is dominated by the Balmer lines (tracing young hot O-B type stars, but also contaminated by older A-type stars) and through the CaII triplet (CaT) at λ~8,500Å (which, in the case of recent star formation, is dominated by young Red Supergiants and young AGB stars, although the latter are also contributed by older stellar populations). Moreover, we have detected (only in the regions of highest signal-to-noise ratio) the weak absorption feature of HeI λ=4,922 (Fig.4a), which is an unambiguous tracer of B-type stars and of stellar populations with age of about 10 Myr[26]. Fig.4b shows the kinematics of the various components as a function of galactocentric distance along the X-shooter slit. Except for the central kpc, where stellar features are probably dominated by the central starburst in the host galaxy, all stellar features in the outflow region, at a galactocentric radius of about 1-3 kpc, show a blueshift relative to the galactic disk (see Methods for the identification of the latter), indicating that they are indeed associated with the outflow. The velocity of the stellar features (reaching a maximum blueshift of about 100 km s$^{-1}$) is much lower than the velocity observed for the nebular lines in the outflow (about 250-450 km s$^{-1}$); however, this difference is expected in the scenario in which stars form in the outflow.

Indeed, although the gas in the outflow is probably maintained at high velocity (at least in the central 1.5 kpc) by the continuous action of radiation pressure or ram pressure by expanding hot gas, as soon as stars form in the outflow they react only to gravity, and are rapidly decelerated by the galaxy gravitation field (that is, they move ballistically). A simple dynamical model (see Methods) can easily describe such an effect. In Fig.4b the solid curve shows the position-velocity track of stars formed at the base of the outflow (specifically at 500 pc from the nucleus, corresponding to a projected distance of about 350 pc), where most of the star formation in the outflow is occurring, as inferred from the broad Hα flux map. Within this model such stars are rapidly decelerated and their velocities even change sign (that is, they fall back towards the disc and towards the bulge), reproducing the *positive* velocity of young stars observed at 5 kpc from the nucleus. Within this model, stars formed in the outflow at projected distances larger than 1 kpc are only mildly decelerated and become gravitationally unbound (e.g. dashed curve in Fig.4b).

Overall, the combined extensive evidence supports the scenario in which stars have formed in the outflow of this galaxy. By using the total Hα emission of the gas in the Eastern (approaching) outflow, and corrected for extinction through the Balmer decrement, we infer a total star formation rate in the Eastern outflow of about 15 M$_\odot$ yr$^{-1}$, assuming a "Chabrier" Initial Mass Function[27].

Performing a similar analysis in the western (receding) outflow is prevented by the large dust extinction. However, if the Western outflow is also characterized by similar star formation, then the total star formation rate occurring in the outflow is about 30 $M_\odot$ yr$^{-1}$, that is, a significant fraction (about 25%) of the global star formation rate of the merging system[28], about 115 $M_\odot$ yr$^{-1}$.

**Acknowledgments:** R.M., S.C. and F.B. acknowledge support by the Science and Technology Facilities Council (STFC). R.M. acknowledges ERC Advanced Grant 695671 "QUENCH". HRR and ACF acknowledge ERC Advanced Grant 340442.

**Authors Contributions** R.M. project led, data analysis and interpretation. H.R. X-shooter data reduction. A.F. and W.I. theoretical modelling. S.C., S.A. and E.B. reduction and analysis of the MUSE data. S.C. and R.G. stellar continuum subtraction and continuum analysis. L.C. interpretation of the near-IR spectra. E.O. nebular and stellar lines identification and diagnostics. F.M., A.M., G.C. and E.S. contribution to interpretation. F.B. comparison with SDSS data.

**Author Information** Reprints and permissions information is available at www.nature.com/reprints. The authors declare no competing financial interests. Readers are welcome to comment on the online version of the paper. Correspondence and requests for materials should be addressed to R.M. (r.maiolino@mrao.cam.ac.uk)

# Figures

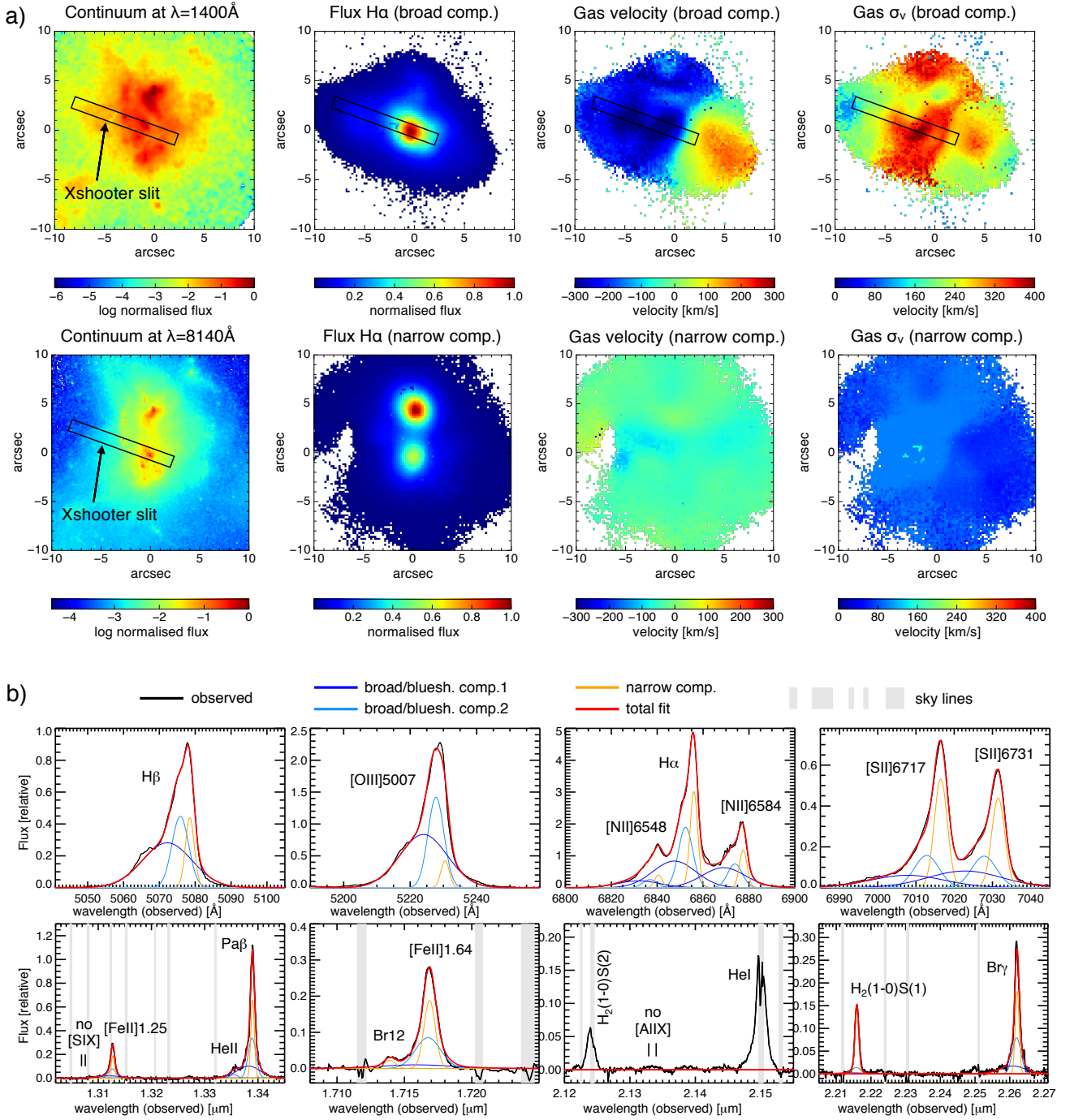

**Figure 1**. **Spectral decomposition of the outflow in IRAS23128-5919. a:** HST continuum images, surface brightness, velocity and velocity dispersion $\sigma_V$ maps of the broad and narrow components of the H$\alpha$ line, inferred from the MUSE data. The location and orientation of the X-shooter slit is shown (boxes). **b:** Subsections of continuum-subtracted X-shooter spectra, extracted from the central region, around some of the relevant emission lines, showing the decomposition between narrow and broad components, as well as the non detection of coronal lines (see Methods).

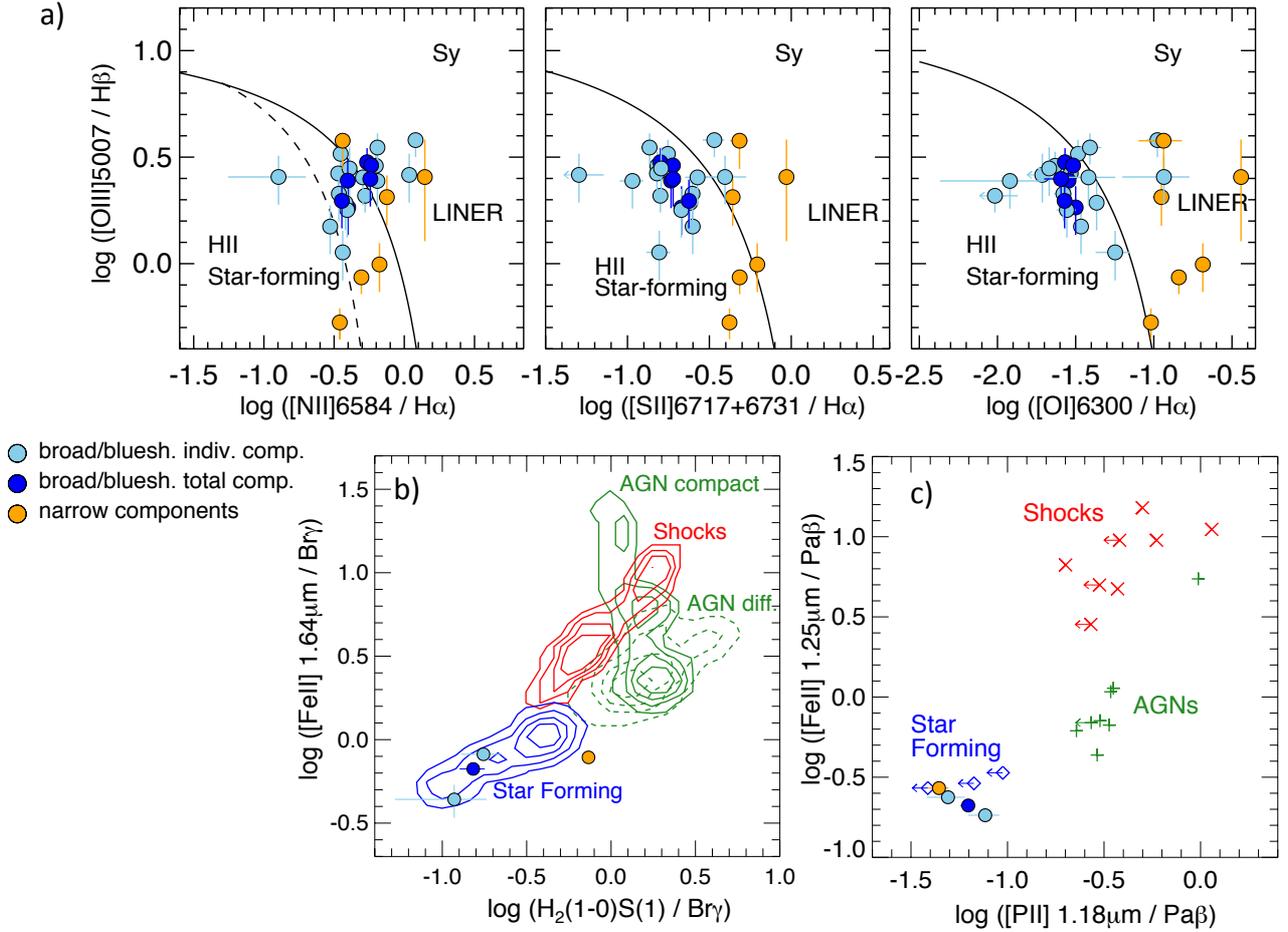

**Figure 2. Diagnostic diagrams.** Distribution of the individual broad Gaussians (light-blue symbols) and of the resulting total broad components (dark-blue symbols) tracing the outflow, as well as of the narrow components (orange symbols) tracing gas in the host galaxy, extracted from different apertures along the X-shooter slit. **a:** BPT diagnostic diagrams involving optical nebular lines, showing the dividing curves between HII star-forming regions, Seyferts, and low-ionization nuclear emitting regions (LINERs)[20,29]. **b:** Diagnostic diagram involving near-infrared nebular lines showing the comparison with the distribution of star forming galaxies, shocked regions, AGN compact regions and AGN diffuse emission[24]. **c:** Diagnostic diagram based on other infrared emission lines, also showing the comparison with the distribution of various classes of sources. Error bars show the 1σ uncertainties.

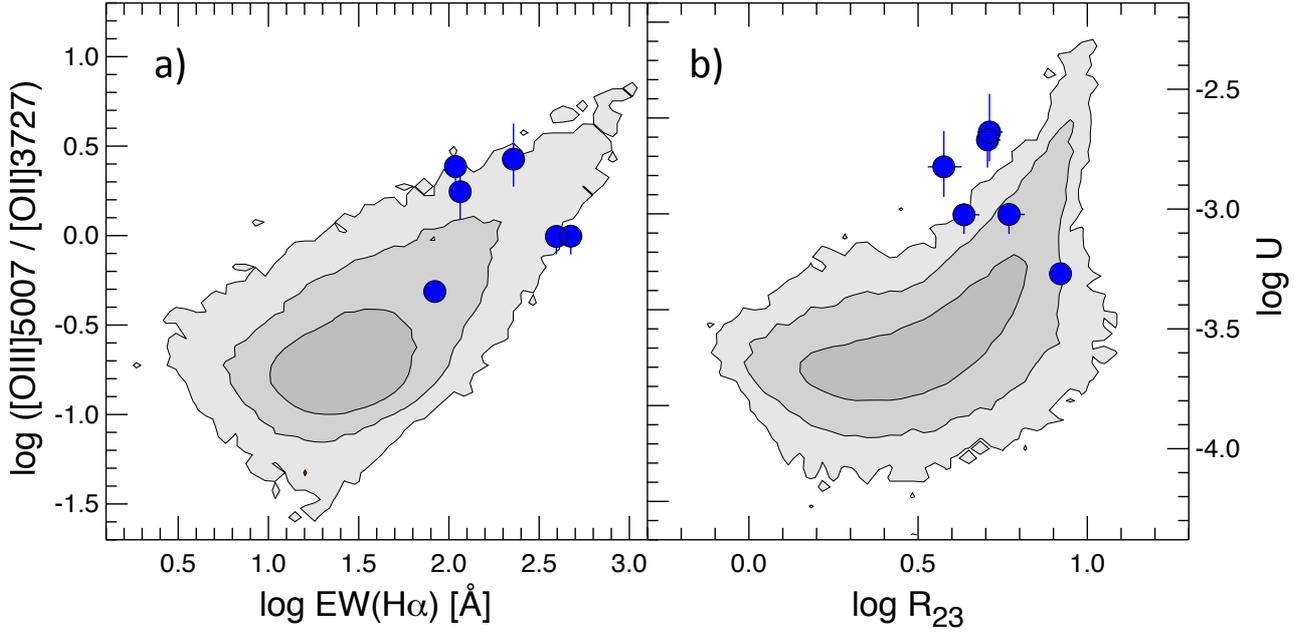

**Figure 3. Ionization parameter.** The [OIII]/[OII] ratio (left-hand y-axis) is sensitive to the ionization parameter U, as given on the right-hand y-axis[30] (although it also has secondary dependence on metallicity). **a:** [OIII]/[OII] as a function of EW(Hα), which is used as a proxy of the age of the stellar population. Contours indicate the distribution of star forming galaxies from the SDSS survey (68%, 95% and 99.7% of the population), illustrating that younger systems have higher ionization potential. Symbols indicate the location of the gas in the Eastern outflow (we cannot differentiate between individual broad components since we cannot differentiate the continuum associated with each component), indicating that it is consistent with *in situ* photoionization by young stars. **b:** [OIII]/[OII] ratio has a function of R23=([OIII]5007,4940+[OII]3727)/Hβ, which is sensitive to the metallicity, with a secondary dependence on ionization parameter. The diagram further supports the idea that the gas excitation in the outflow does not differ substantially from normal star forming regions and, if anything, the ionization parameter is even higher. Error bars show the 1σ uncertainties.

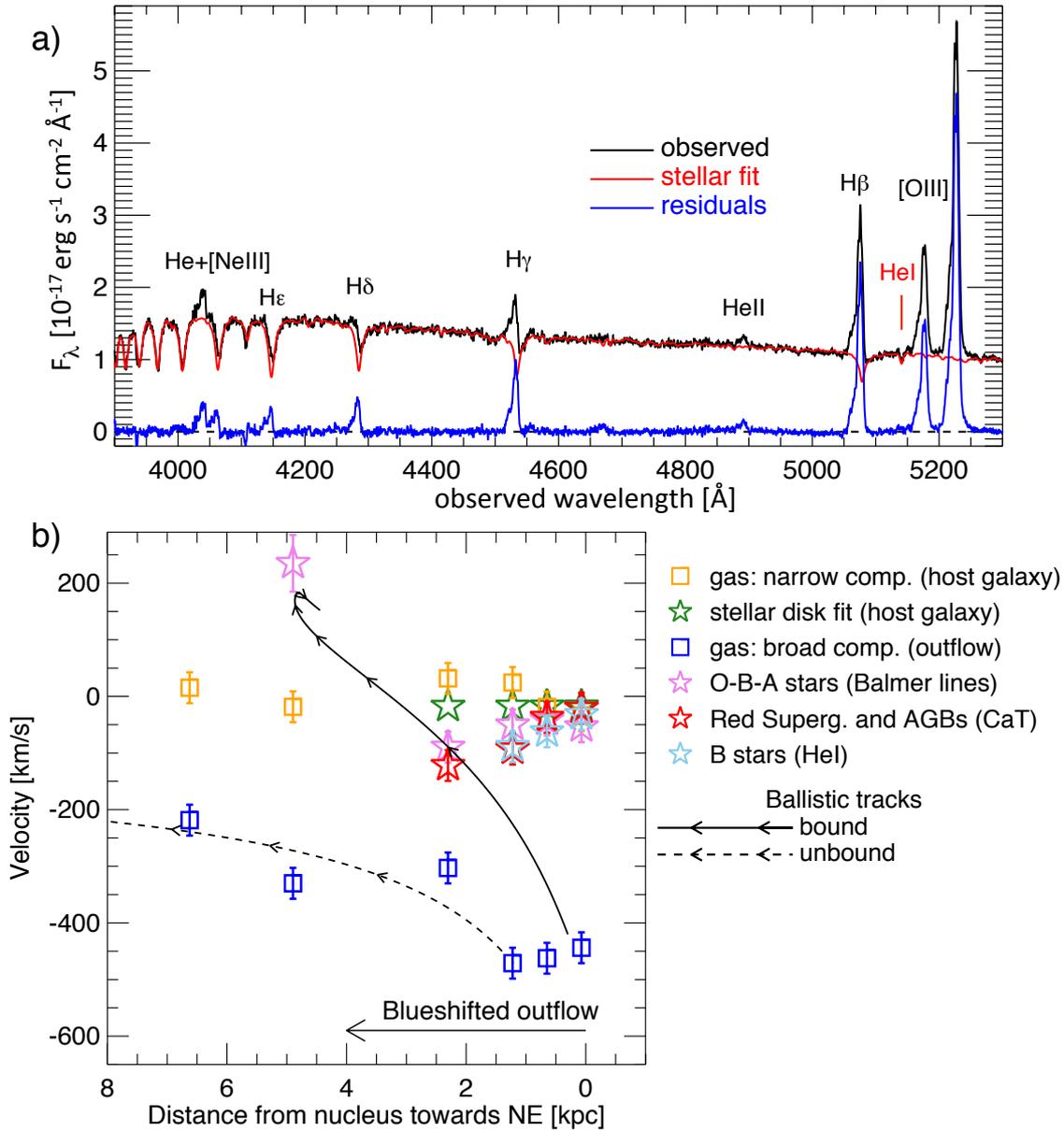

**Figure 4. Kinematics of young stars compared with the gas kinematics. a:** Stellar fit of the spectrum extracted from the outflow region at 1kpc from the nucleus. **b:** Velocity of the young stellar population along the outflow, traced by different features (see also the Methods) compared with the kinematics of the broad nebular component tracing the gaseous outflow (centroid of the broad Gaussians fitting the broad component; dark blue symbols), and the kinematics of the host galaxy disc traced by the narrow nebular component (orange symbols) and by a fit of the disc stellar kinematics in the MUSE data (green symbols, see Methods). Black curves show the position-velocity tracks of stars formed in the outflow, according to the model described in the Methods. Error bars show the 1σ uncertainties.

**Methods**

1. **Data reduction and analysis**

The optical integral field spectroscopic data were retrieved from the ESO archive and were obtained with the MUSE spectrometer at the ESO-VLT with an on-source integration time of 52 minutes. The wavelength coverage is from 0.46 to 0.93μm and the spectral resolution varies from R=2000 to R=4000 across the wavelength range. The seeing during the observations was only about 2.5". The stellar continuum was fitted and subtracted at each pixel by using the PPXF routine[31] by using the MILES templates[32]. The nebular lines Hα, [NII]6548, [NII]6584, [SII]6717, [SII]6731, [OI]6300, Hβ, [OIII]4959, [OIII]5007 were simultaneously fitted with two Gaussian components, a narrow one with FWHM<300 km/s and a broad one with FWHM>300 km/s. At each spatial position the velocity and velocity dispersion of each of the two components was fixed for all nebular lines. The intensities of the [NII] doublet and of the [OIII] doublet were tied to have the ratio given by their Einstein coefficients. The resulting maps of the Hα flux, velocity and velocity dispersion of the two components in Fig.1a are obtained by masking out regions for which the signal-to-noise of the component is lower than five. It shall be noted that in previous works[16], which analyzed VIMOS data of the same source, the most eastern nebular emission was associated with the "narrow" component but fitted with very large velocity dispersion ($\sigma_V$~300 km/s) and very blueshifted (velocity v< −300 km/s). This mismatch was likely due to the lower signal-to-noise of the VIMOS data in those external regions, which made the line decomposition much more difficult. In our analysis, using MUSE data, we have corrected this (not completely appropriate) association: such high values of the velocity dispersion and strong blueshift should be actually associated with the "broad" blueshifted component which traces the outflow.

The new spectra were obtained with the spectrometer X-shooter[33] at the ESO-VLT. We selected the 11" long slit with a width of 1.3" in the UVB arm (wavelength spectral range 300-560 nm), 1.2" in the VIS arm (spectral range 550-1020 nm) and 1.2" in the NIR arm (spectral range 1020-2480 nm). The resulting spectral resolutions are R=4000 in the UVB, R=6700 in the VIS and R=3890 in the NIR. The seeing during the observations was about 0.9" on average. The slit was positioned with a Position Angle of 70° with the southern nucleus centered at 3.2" from the centre of the slit (Fig.1a); this orientation and centering enables the slit to sample both the nuclear and the eastern part of the outflow. Observations were executed by nodding the slit by a few arcsec (to optimize the removal of detector artifacts) and interleaved with blank sky exposures, obtained at about 1 arcmin from the galaxy, for sampling the background. The total on-source integration time was 1.7 hours. Data reduction and calibration was performed following the standard pipeline steps[34]. Great care was taken to correct the atmospheric absorption features, not only in the near-IR, but also in the optical, by using the telluric standard and by smoothing its spectrum to the resolution of the science observations (the spectral resolution of the telluric standard is higher, by about 30%, then the science observation because the seeing was smaller than the slit).

We extracted six spectra along the slit, with variable apertures (from 0.5" to 2.5") to find a tradeoff between spatial information and signal-to-noise. The continuum spectra were fitted with PPXF[31] by using the MILES stellar synthesis library[32]. The whole library was used in the fitting. However, it should be noted that unfortunately, templates covering the required spectral range at the high resolution of our spectra are available only for ages down to 30 Myr. The latter constraint implies that we cannot disentangle populations younger than 30 Myr. More specifically, all spectral fits require a population of 30 Myr (jointly with contribution from older stellar populations), but this shall be regarded as an upper limit of the youngest stellar population because of the library limits. We adopted two strategies for fitting the continuum: 1) we masked all of the nebular lines (over their whole width, about 30-40Å) and fitted the remaining portion of the spectrum; 2) we fitted simultaneously the nebular lines (with their multiple velocity components) and the stellar continuum without any masking. The resulting fluxes and decomposition of the nebular lines do not change depending on the method. However, based on simulations, the stellar kinematics (especially the young component associated with the Balmer lines) is better recovered by the second method. The HeI emission+absorption line was fitted separately, as this feature may trace a different stellar population with a different kinematics. For the latter fitting the shape of the HeI emission was fixed to that of [OIII]5007, but allowed to vary in flux. Regarding the fitting of the CaII triplet (which was fitted independently of the rest of the spectrum), we have restricted it to the bluest of the three lines ($\lambda$=8498Å) since the other two lines are heavily affected by sky emission lines and telluric absorption.

The different stellar populations in the stellar fit should in principle be allowed to have different velocities. In particular, we should in principle allow the old stellar population to have a different velocity relative to the young stellar population. However, this is not really possible in practice because: 1) this additional degree of freedom would introduce a lot of degeneracy when also attempting to simultaneously deblend the nebular line emission from the stellar features, and 2) the young stellar population dominates the observed stellar light across the entire spectrum, hence the velocity traced by the spectral features associated with the old stellar population is extremely difficult to recover.

Note that the signal-to-noise in the spectra allows us to trace the stellar kinematics through the HeI line only out to 1.2 kpc, and through the CaII triplet only out to 2.3 kpc. At 5 kpc the stellar kinematics can be recovered only through the Balmer absorption lines. At 6.6 kpc (outermost point at which the spectrum is extracted) the very low signal-to-noise on the continuum does not allow us to detect stellar signatures.

It is interesting to note that the spectrum extracted at 2.3 kpc (sampling from 2" to 3.6" from the nucleus) includes a clear knot observed in the HST image. This could be a star cluster that has formed within the outflow. Indeed, in the position-velocity diagram (Fig.4b) it is located on the ballistic track of particles formed in the inner part of the outflow.

Regarding the specific fitting of the nebular lines, in the optical spectra, the higher spectral resolution and higher signal-to-noise of the X-shooter data reveals that the broad blueshifted component, tracing the outflow cannot be reproduced with a single Gaussian. In the central aperture the broad component requires at least two blueshifted broad Gaussians (blue lines in Fig.1b) to be properly fitted (along with a narrow component, accounting for gas in the host galaxy). In some of the eastern apertures, dominated by the outflowing gas, the broad blueshifted component requires three Gaussian components to be properly fitted. In the latter apertures a very weak narrow component (even more clearly separated from the broad component), associated with the outskirts of the host galaxy in the background is also always detected.

The fitting of these components in the X-shooter spectra is performed by imposing that the center and width of each of these components is the same for all nebular lines and only their amplitude is allowed to vary. For the two nitrogen lines [NII]6548,6584 and the two oxygen lines [OIII]4959,5007 their relative intensities was fixed to match the relative values of the Einstein transition coefficients.

In the central aperture, absorption by a Diffuse Interstellar Band at 6282Å affects the blue shoulder of the bluest and broadest component of the [OI]6300 line, so the spectral region in the range 6125Å<$\lambda_{rest}$<6291Å was not included in the fit. This results into a slightly higher uncertainty of the fit parameters for this component of the [OI] line.

For the [OII]3727 doublet there is some degeneracy on the fit of the three components tracing the outflow in the eastern spectrum. However, since we do *not* use the relative intensity of the two lines of the [OII] doublet to infer information on the gas density, but we only use their sum to investigate the ionization parameter (along with [OIII]), such degeneracy does not affect our results.

The line profiles (e.g. Fig.1b) have clear humps and changes of slope, which leave little ambiguity about the multi-component decomposition of the lines. However, since ascribing a specific physical meaning to each individual broad component is not relevant for this paper, we also combine all broad components together in the diagnostic diagrams of Fig.2. This may still leave some ambiguity about the decomposition of the broad and narrow components. However except for the nuclear region, in most of the outflow the line flux is totally dominated by the broad component, hence even if the narrow component were totally incorporated into the broad component this would not change the location on the BPT diagrams of Fig.2.

For the optical nebular lines correction for dust reddening (by using a Milky Way extinction curve[28]) was done by using the Hβ/Hγ ratio to infer the reddening. Hα/Hβ cannot be used as these two lines were observed with different apertures.

In the near-IR data of the nuclear region (which is the only region where we can extract the spectrum with a signal-to-noise high enough for a proper spectral decomposition), the profile of the nebular lines changes substantially relative to the optical lines. This is primarily due to the effect of differential dust extinction, as a consequence of which some components are much more absorbed in the optical than in the near-IR. Therefore, we have fitted separately the near-IR lines with a different set of components, not tied to the parameters of the optical components, although the general properties are similar.

In the near-IR the diagnostic line ratios are corrected for dust reddening by using a Milky Way extinction curve[38] and by using the Brγ/Paβ ratio to infer the reddening (although the inferred reddening is low and most of these line ratios are actually unaffected by dust reddening).

Some additional near-IR spectral region of interest, which were not shown in the main paper, are shown in Extended Data Figure 1.

We note that in many cases (especially thanks to the high signal-to-noise of the optical spectrum) the formal errors on the central velocity is very small, both for the gaseous and the stellar features. However, in Fig. 4b we conservatively give a minimum error bar corresponding to half of the spectral resolution element.

## 2. Near-IR diagnostic diagrams

As the near-IR diagnostic diagrams are generally less used (due to the greater technical/observational difficulties in obtaining near-IR spectra with appropriate sensitivity) in this section we provide some additional details about their physical meaning and about the way they were obtained.

The near-IR iron lines are extremely useful to identify shock excitation. Indeed, shocks are known to destroy dust grains, hence releasing into the ISM large amount of iron, which is otherwise locked into grains. This results in the emission of iron near-IR transitions, relative to the hydrogen recombination lines, being much larger than typically observed in star forming regions. Shock heating of the ISM also results into excitation of the near-IR vibrational transition of molecular hydrogen. These are the reasons why the [FeII]1.64μm/Brγ versus H$_2$(1-0)S(1)2.12μm/Brγ diagram (Fig.2b) is so effective in disentangling shocks from star forming regions, as outlined by Colina et al. (2015)[24], from which the distribution of the different classes of excitation sources have been taken.

The low ionization line [PII]1.18μm is an excellent diagnostic of partially ionized, transition regions produced by X-rays, either produced by radiative shocks or by AGNs[25]. The ratios [FeII]1.25μm/Paβ and [PII]1.18μm/Paβ shown in Fig.2c for different classes of objects have been obtained from the literature[25,35,36,37] and illustrate that this diagram is also very effective in disentangling different excitation mechanisms. Note that the comparison with AGNs is limited to the "type 2" class to avoid contribution to Brγ and Paβ from the Broad Line Region.

3. Coronal lines

Coronal lines are not detected in our X-shooter spectrum of IRAS2318-59. In particular, none of the coronal lines observable in good regions of atmospheric transmission ([SVIII]0.9913μm, [FeXIII]1.0747μm, [SIX]1.252μm, [AlIX]2.045μm) is detected. Although coronal lines are not always seen in AGNs, most of the powerful AGN do show evidence for coronal lines in the near-IR (which is the wavelength range hosting many of the most intense coronal lines), especially if observed with appropriate signal-to-noise and for those AGNs not suffering dilution from star formation in the host galaxy[35,39,40,41,42,43,44]. For instance, in the prototypical Sy2 galaxy Ciricinus, [AlIX]2.045μm is three times stronger than then nearby HeI2.058μm line[42], while in IRAS2318-59 [AlIX]2.045μm is completely undetected (despite being in a region of good atmospheric transmission and unaffected by OH sky lines), implying that it is at least 20 times *fainter* than the HeI2.058μm line (Fig.1b).

However, previous works[44] have claimed the detection of weak [SiVI]1.96μm emission in IRAS2318-59. This line is shifted in a bad region of atmospheric transmission and also plagued by a prominent OH sky line, so it is difficult to assess its presence even with the high S/N of our spectra. Extended Data Fig.1b shows the nuclear spectrum of IRAS2318-59 extracted from the nuclear aperture, zoomed around the expected location of [SiVI]. The vertical lines indicate the expected location of the three Gaussian components used to fit the nebular lines in the near-IR spectrum. While no features are associated with the narrow and intermediate broad components, there is indeed a weak feature around the expected location of the broadest component. However, the detected feature is much narrower than the expected component at this wavelength. Indeed, if included in the simultaneous fit together with all the other nebular IR lines, the fitting procedure sets to zero the flux of the putative [SiVI] component associated with the observed feature, because it cannot accommodate the narrow profile of the feature. It could be that this feature is actually highly ionized gas ionized by the AGN in the central region of IRAS2318-59 and which has kinematic properties different from all the other nebular lines. Alternatively, it should be noted that the feature is located exactly at the wavelength of one of the deepest atmospheric features in the K-band, which is very difficult to correct, so the observed feature could be a residual of an imperfect correction of the telluric absorption. However, even if the feature is real and associated with [SiVI] this would simply confirm the presence of the obscured AGN in the southern nucleus of IRAS2318-59, but it would not affect our conclusion that the photoionization of the bulk of the gas associated with the broad blueshifted component tracing the outflow is not due to the AGN, both because the observed coronal line is very weak (a factor of 30 weaker than observed in other classical AGNs such as Circinus and NGC1068, relative to Brγ) and because appears to have different kinematic properties.

## 4. HeII line detection

It is interesting to note that the HeII4686 line is clearly detected in the outflow, although weak (Fig.4a). Its intensity relative to Hβ ($F_{HeII4686}/F_{H_\beta} \sim 0.07$) is much lower than observed in the Narrow Line Region of AGNs (for which generally $F_{HeII4686}/F_{H_\beta} \sim 0.3$-$0.5$)[29,43], but it is close to the value observed in very young star forming galaxies ($F_{HeII4686}/F_{H_\beta} \sim 0.01$-$0.08$)[45,46]. In young star-forming galaxies the weak HeII emission is ascribed to excitation by radiative shocks associated with SNe or stellar winds, which contribute little to the overall ionization budget but can produce enough hard photons to generate some highly ionized species[45,46]. In addition, HeII can also be excited by the contribution of Wolf-Rayet stars, associated with very young stellar populations (a few Myr old), which would be in line with the other observational evidence for a young stellar population in the wind.

## 5. Dynamical model

We first clarify that the model shown in Fig.4b (solid and dashed lines) is not meant by any means to be the unique model that can fit the data, neither the best model. The limited information available does not allow us to discriminate among different dynamical models and between different set of parameters within such models. The simple model shown in Fig.4b is only meant to show that it is possible to explain the kinematics observed for the stars in a simple scenario in which stars are formed within the outflow (with an initial velocity corresponding to the gas in the outflow at that location) and then move ballistically within the galaxy gravitational potential, making them decelerate and even fall back onto the galaxy disc/bulge. This section is aimed at providing some more details about the simple model specifically used in Fig.4b.

The gravitational potential within the central few kpc of the galaxy is supposed to follow the axisymmetric Miyamoto-Nagai disc+bulge in the form:

$$\Phi(r,z) = -\frac{GM}{\sqrt{r^2 + \left(a + \sqrt{z^2 + b^2}\right)^2}}$$

where $M$ is the total dynamical mass of the system, G is the gravitational constant, $r$ is the radial cylindrical coordinate on the plane of the disc, $z$ is the coordinate perpendicular to the disc plane, and $a$ and $b$ are constants. The curves shown in Fig.4b assume $M=1.5\ 10^{11}$ $M_\odot$, $a=1.7$ kpc and $b/a=0.1$. The dynamical mass may appear somewhat large, and in particular larger than given in previous works[16]; however, one should take into account that 1) in the Miyamoto-Nagai potential the quantity $M$ is the total mass (integrated to infinity) associated with the potential and 2) velocity projection effects make it difficult to infer a solid dynamical mass of the system, especially for nearly face-on systems (in particular the inferred dynamical mass is generally a lower limit, modulo the cosine of the inclination angle). For sake of simplicity and for sake of reducing the number of free parameters we have assumed that the galactic disc is oriented perpendicular to the line of sight (this is a reasonable assumption given that the rotation curve of the disk is very weak). In the model the velocity of the outflowing gas makes an angle of ~45° with the line of sight. The solid curve corresponds to particles (stars) starting from 500pc from the nucleus (corresponding to a projected distance of about 350pc), while the dashed curve corresponds to particles (stars) starting from distance of 2kpc from the nucleus (corresponding to a projected distance of 1.4kpc). We emphasize, once again, that there is large degeneracy between all of these parameters and that different combination of these parameters (as well as different gravitational potential) can reproduce the observed stellar kinematics equally well. The location of the arrows along the curves in Fig.4b indicate time intervals separated by 5 Myr.

## 6. Host galaxy disc reference velocity field

It is important to identify the reference velocity field of the (dynamically quiescent) disc of the host galaxy. As in many other studies[14,16,18,47], the narrow component of the nebular emission lines is tracing gas in the galaxy disc. The velocity field traced by the narrow component is shown in Fig.1a and by the orange symbols in Fig.4b, illustrating that it is relatively flat (disc likely close to face-on).

It would be useful to have also a reference for the kinematics of the stellar population in the stellar disc. This is more difficult to trace, as in this paper we have shown that young stars are formed in the outflow and, therefore, the stellar kinematic component is likely dominated by the outflowing stars in several regions. In particular, we have discussed that the stellar population traced by the Eastern part of the X-shooter slit is likely dominated by stars formed in the outflow. We have attempted to trace the stellar kinematics in other regions by exploiting the MUSE cube. This is quite difficult since both the signal-to-noise and the spectral resolution of the MUSE spectra are much lower than in the X-shooter spectra. Moreover, the seeing during the MUSE observations was as bad as ~2"; the latter is a major issue for investigating the spatially resolved properties of the stellar population, since it implies that the strong emission from the centrally concentrated stellar population heavily contaminates the stellar light observed in the outer, fainter stellar disk, as a consequence of the wide wings of the Point Spread Function.

Bearing in mind these various major issues, we have attempted to recover the velocity field of stars possibly not associated with the outflow and likely associated with the stellar disc, as discussed in the following. We have first undertaken a Voronoi binning of the MUSE cube to reach a signal-to-noise higher than 35 per spectral pixel at the wavelength of the CaII triplet. The resulting continuum flux with such a rebinning is shown in Extended Data Fig.2a. Then we have performed a PPXF fitting of the reddest line of the CaT (which is the least affected by telluric absorption) and we have derived the velocity of the stellar population, which is shown in Extended Data Fig.2b. Clearly, the stellar velocity field traced by MUSE is noisier than the X-shooter spectrum because of the issues discussed above. Moreover, it is also clear that the MUSE data do not recover the full kinematic information obtained from the X-shooter spectra; in particular, the velocity of the stars do not reaches the blueshifted values observed in X-shooter; this is clearly a consequence of the poor quality of the MUSE data (much worse seeing, Extended Data Fig.2b, poorer spectra resolution and lower signal-to-noise). Yet, the 2D stellar kinematics maps shows that the southern galaxy has a hint of rotation. The region towards the Northeast of the southern nucleus shows an excess of blueshifted stellar velocities, in rough agreement with the X-shooter data, modulo the higher noise and the beam smearing issues discussed above. Within all the uncertainties discussed above, the observed kinematics can be interpreted as a mild (face-on) rotation curve, with an outflowing stellar component superimposed towards the Northeast, and everything smeared by the large seeing. We have attempted to model the rotational component by masking the northern galaxy (shaded out in

Extended Data Figure 2), by also masking the region towards the Northeast that is probably affected by the outflow and by also convolving the model with the large seeing of the MUSE observation. The resulting model is shown in Extended Data Fig.2c.
The model should provide an indication of the stellar rotation field beneath the outflowing stars. We have extracted the inferred stellar rotation field at the same points of extraction of the X-shooter spectra. The resulting velocities are shown with green stars in Fig.4b.

## 7. Luminosity of the Hα broad component in the outflow

Determining the absolute luminosity of the broad component of Hα is important to infer some key properties of the outflow. We focus on the regions in which the outflow is approaching (i.e. the blueshifted broad component), as the receding component is more heavily affected by dust reddening, hence affected by larger uncertainties. By selecting, in the MUSE cube, those regions where the broad component is blueshifted by more than 150 km/s, and by correcting the line flux (in each spaxel) by dust reddening as inferred by the Hα/Hβ Balmer decrement, the inferred broad line luminosity is $2.8 \cdot 10^{42}$ erg s$^{-1}$. If this luminosity is used as an indicator of ionizing photons, this can be translated into a star formation rate[27] of 15 M$_\odot$/yr.
By assuming a typical gas density, this luminosity can also be used to infer the mass of the warm ionized component of the outflowing gas. More specifically, Hα luminosity and mass of ionized warm gas are related by the following relation[48]

$$M_{gas} = 6.1 \cdot 10^8 M_\odot \left(\frac{L_{H\alpha}}{10^{44}\ erg/s}\right)\left(\frac{\langle n_e \rangle}{500\ cm^{-3}}\right)^{-1}$$

where $\langle n_e \rangle$ is the average electron density in the warm ionized clouds emitting Hα (and where we have neglected the clouds density contrast factor). By assuming a gas density of 500 cm$^{-3}$, the observed luminosity of the Hα broad component implies a mass of ionized gas in the outflow of $1.7 \cdot 10^7$ M$_\odot$. However, this is a very conservative lower limit, since only a small fraction of the gas is in the warm ionized phase, while various studies have shown that the bulk of the mass is in the molecular and atomic neutral phases. According to some works[48] the molecular phase in galactic outflows can be two orders of magnitude higher than the ionized phase. This would imply that the gas mass in the outflow of IRAS2318-59 could easily be higher than $10^9$ M$_\odot$. Millimeter observations of molecular transitions are needed to properly assess the total gas mass in the outflow.
We note that the quantities inferred above only refer to the approaching component of the outflow. If the receding side has similar properties, then the star formation rate and gas mass in the whole outflow should be doubled.

**Data Availability Statement**
The X-shooter observations presented in this paper were obtained through the ESO progamme 293.B-5032 and the associated data are available at the ESO archive: http://archive.eso.org/cms/data-portal.html. The reduced HST, MUSE and X-shooter data that support the findings of this study (and used to produce all figures) are available from the corresponding author upon reasonable request.

**References for the Methods:**

31. Cappellari, M. Improving the full spectrum fitting method: accurate convolution with Gauss-Hermite functions. arXiv:1607.08538 (2016)

**Extended data**

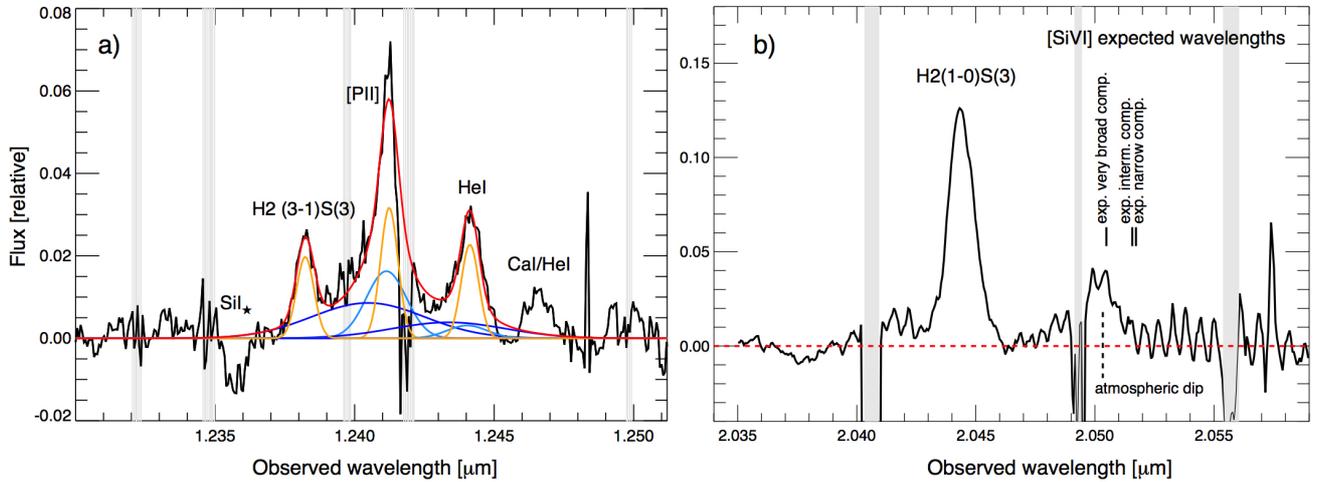

**Extended Data Fig.1. Additional sections of the near-IR nuclear spectrum. a:** Spectrum around the [PII]1.188mm line. The fitting components have the same color coding as in Fig.1b. **b:** Spectrum around the expected wavelength of [SiV]1.96μm. The expected location of the broad and narrow components of [SiVI] is marked. A feature corresponding to the expected location of the broadest component is observed, but is much narrower than the width of the same component observed in other nebular lines, and is at the location of an atmospheric absorption dip.

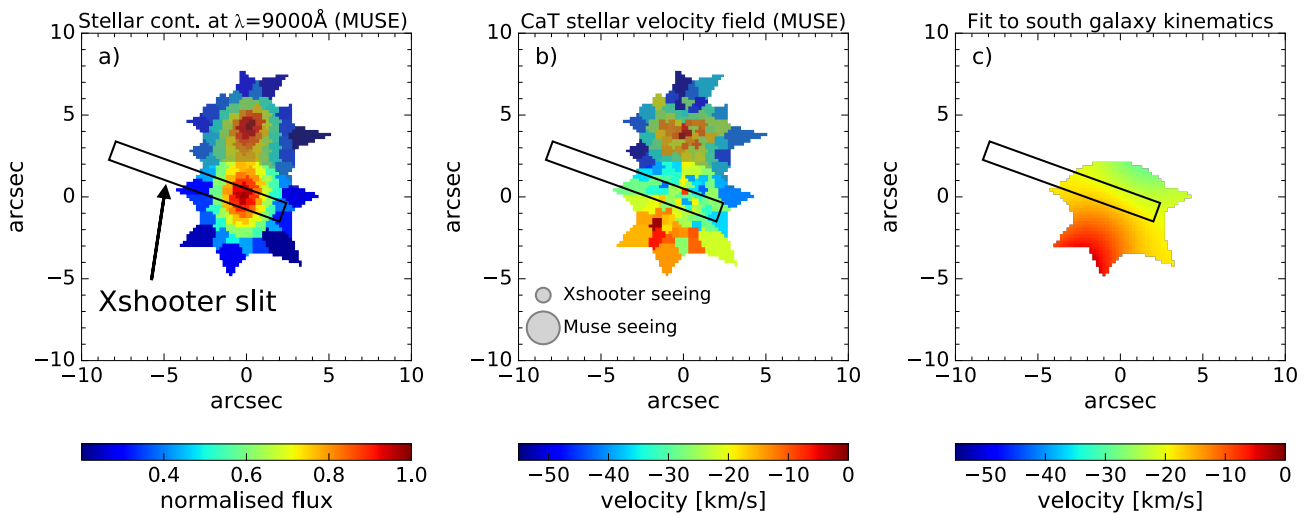

**Extended Data Fig.2. Stellar velocity field from the MUSE data.** Although the quality of the MUSE data is not adequate to extract a very reliable stellar velocity field, panels **a)** and **b)** show the distribution of the stellar continuum around λ=9000Å and the velocity field inferred from the reddest line of the Calcium Triplet (CaT), by applying Voronoi binning to the MUSE cube. Panel b) also shows the seeing difference between the X-shooter observation and the MUSE observation. Panel **c)**

shows a rotation velocity fit to the southern galaxy, and by masking the region to the Northeast (around the X-shooter slit), which is probably affected by outflowing stars.